\documentclass[sigconf,manuscript]{acmart}

\usepackage{booktabs}
\usepackage{multirow}

\AtBeginDocument{%
  \providecommand\BibTeX{{%
    \normalfont B\kern-0.5em{\scshape i\kern-0.25em b}\kern-0.8em\TeX}}}

\setcopyright{acmcopyright}
\copyrightyear{2018}
\acmYear{2018}
\acmDOI{XXXXXXX.XXXXXXX}

\acmConference[Conference acronym 'XX]{Make sure to enter the correct
  conference title from your rights confirmation emai}{June 03--05,
  2018}{Woodstock, NY}
\acmBooktitle{Woodstock '18: ACM Symposium on Neural Gaze Detection,
 June 03--05, 2018, Woodstock, NY} 
\acmPrice{15.00}
\acmISBN{978-1-4503-XXXX-X/18/06}

\begin{document}

\title{The Grind for Good Data: Understanding ML Practitioners' Struggles and Aspirations in Making Good Data}
\renewcommand{\shorttitle}{Understanding ML Practitioners' Struggles and Aspirations in Making Good Data}

\def\thefootnote{}\footnotetext{* These authors contributed equally to this work.}

\author{Inha Cha$^{*}$\thefootnote{}}
\email{inhacha@upstage.ai}
\affiliation{%
  \institution{Upstage AI Research}
  \city{Yongin}
  \country{Korea, Republic of}
}

\author{Juhyun Oh$^{*}$\thefootnote{}}
\email{juhyun@upstage.ai}
\affiliation{%
  \institution{Upstage AI Research}
  \city{Yongin}
  \country{Korea, Republic of}
}

\author{Cheul Young Park$^{*}$\thefootnote{}}
\email{cheul@upstage.ai}
\affiliation{%
  \institution{Upstage AI Research}
  \city{Yongin}
  \country{Korea, Republic of}
}

\author{Jiyoon Han}
\email{jiyoonhan@upstage.ai}
\affiliation{%
  \institution{Upstage AI Research}
  \city{Yongin}
  \country{Korea, Republic of}
}
 
\author{Hwalsuk Lee}
\email{hwalsuk.lee@upstage.ai}
\affiliation{%
  \institution{Upstage AI Research}
  \city{Yongin}
  \country{Korea, Republic of}
}

\renewcommand{\shortauthors}{Cha, Oh and Park, et al.}


\begin{abstract}

We thought data to be simply given, but reality tells otherwise; it is costly, situation-dependent, and muddled with dilemmas, constantly requiring human intervention. The ML community’s focus on quality data is increasing in the same vein, as good data is vital for successful ML systems. Nonetheless, few works have investigated the dataset builders and the specifics of what they do and struggle to make good data. In this study, through semi-structured interviews with 19 ML experts, we present what humans actually do and consider in each step of the data construction pipeline. We further organize their struggles under three themes: 1) trade-offs from real-world constraints; 2) harmonizing assorted data workers for consistency; 3) the necessity of human intuition and tacit knowledge for processing data. Finally, we discuss why such struggles are inevitable for good data and what practitioners aspire, toward providing systematic support for data works.
\end{abstract}



\maketitle

\section{Introduction}

The value of data was long known to the Machine Learning (ML) community, only to be rediscovered recently. From lauding the effectiveness of large data \cite{halevy2009unreasonable}, to recognizing their limits \cite{Boyd2012-tq, Sun2017-tk}, and now to a new paradigm of ML engineering with data at the center, the data research in ML is at the cusp of change \cite{dcai-web}. As the saying goes, "Garbage in, Garbage out," good data is a binding condition for a successful ML system, but in disappointment to its popularity, it is hard to find. Recent works have eagerly revealed many shortcomings of existing datasets and algorithms derived from them, and to list just a few, they are riddled with biases \cite{Mehrabi2019-vk}, under/mis-represent the real-world \cite{Buolamwini2018-qj, zhao-etal-2018-gender}, outright offensive \cite{crawford2019excavating, Prabhu2020-gk}, exploits spurious correlations \cite{Geirhos2020-ec}, all the while used with little account of by whom and how the data was created \cite{Denton2020-si, Geiger2020-yf}. But indeed, this outpour of work revealing what has been swept under the rug is a blessing in disguise, as at last the ML community is given a chance to redeem itself. The community is striving to this end; existing benchmarks are being fixed \cite{Yang2019-ix,Bowman2021-ok}, more rigorous documentation is emphasized \cite{Gebru2018-xt, Mitchell2018-su, Bender2018-gm}, and new value-centric datasets are introduced \cite{bloom-building, Galvez2021-rz}. More broadly, the movement is pointing towards greater accountability, and transparency in data creation and use \cite{Hutchinson2021-ef}, and a sociocultural system supporting data excellence for ML \cite{Paullada2021-ja, Sambasivan2021-zg, Aroyo2022-tx}.

In this thread, a growing body of work is investigating data through the lens of humans involved in the data work \cite{Pine2022investigating}. Data, from its creation to its utilization in ML or any other application, is naturally human-driven. More so, in this age of deep learning fueled by big data, data is a feat of collective people with variegated skills \cite{Huff2015-qq, Zhang2020How}. Nonetheless, this heterogeneity of humans involved in data work makes organizing data development (creation, preparation, and evaluation altogether) into a structured discipline a convoluted problem. \citet{Karmaker_Santu2021-oz} points out that the need for much human interaction and the subjective nature of data hinder the systematic studies of data construction steps (e.g., data schema development, data annotation) \cite{Amershi_undated-ej} also recognizes that discovering, managing, and versioning the data needed for ML applications is much more complex and difficult than other types of software engineering. In this regard, \citet{Paullada2021-ja} calls for looking into human labor, arbitrary judgments and biases, and unforeseeable circumstances in dataset creation to attain data excellence in ML.

An expanding amount of HCI literature focuses on the human labor that goes into producing, gathering, maintaining, curating, analyzing, deciphering, and disseminating data \cite{Muller2019how, Sambasivan2022all, Thakkar2022when}. Most recently, the CHI 2022 workshop highlighted the need to detail the practices and processes of humans in data work. In line with this trend, we aimed to discover the specific tasks and challenges of humans in data work, with a particular interest in data construction for ML. 

Towards this goal, our study empirically examines what ML practitioners do and consider when creating data. We conducted in-depth interviews focused on the data construction pipeline with 19 ML experts, ranging from a data manager to a senior AI research engineer. We asked our participants about the actual steps they take to build datasets, the difficulties they encounter in the process, and the kind of assistance they deem necessary for constructing good data. Based on our findings, we document a human-oriented data construction pipeline for end-to-end ML systems composed of 6 steps with 14 tasks. We also discover what humans actually consider to make good data in each step of the process. Overall, we discover that individual strategies for constructing datasets are never identical. Nevertheless, our participants share important commonalities in making good data: iteration is the status quo, manual evaluation is inevitable, annotators dictate data quality, and tacit knowledge is the key to improving data to improve a model. 

We also identify three challenges of data construction that repeatedly occur at different stages of the data building process. First, recurring dilemmas result from real-world constraints such as the trade-off between cost versus scalability, consistency, and data validity. Harmonizing diverse data workers involved in data construction to result in consistent and quality outcomes was another. Finally, data construction requires human intuition and tacit knowledge empirically gained through experience to improve data to improve a model in turn. 

Acknowledging the iterative and repeatable characteristics of data work, we discuss future research directions to reduce the trial and error to make good data:

\begin{enumerate}
    \item There is no silver bullet for how to refine data to make a better ML system, but systematic support can help make good data. For example, a model-independent metric for evaluating data quality; aligning qualitative and quantitative metrics of model performance; and possibly explainable model results that allow data debugging.
    
    \item The quality of annotation and the annotators are the most important factors for data quality; thus, we advise providing annotator support: for example, developing an interface that enables efficient annotation; promoting ways to enhance annotators’ understanding of their work; facilitating the annotation task management; and emphasizing the importance of constant and quick communication between annotators and project managers.

\end{enumerate}

Our paper makes three main contributions:

\begin{enumerate}
    \item We document detailed steps and tasks of the data construction pipeline for developing an ML system drawn from semi-structured interviews with 19 ML experts. 

    \item We illuminate the specific human-centric nature that makes structuring data work challenging, paving a stepping stone towards systematizing data work. 

    \item Finally, we suggest the future directions to address the challenges for more productive dataset construction.

\end{enumerate}

\section{Related Works}
\subsection{Data Excellence in ML}
Recently, model-focused ML research has moved into a new phase emphasizing data \cite{Bowman2021-ok, Sun2017-tk, Gebru2018-xt, Mazumder2022-kg}. Data significantly affects the ML system's performance, fairness, robustness, safety, and scalability \citep{halevy2009unreasonable, Sambasivan2021-zg, Mehrabi2019-vk, Zhao2019-br, Wilkinson2020-zg, Dixon2018-vn, Prabhu2020-gk, Yang2019-ix, Buolamwini2018-qj}. However, most popular benchmark datasets in ML applications suffer from pervasive label errors \cite{crawford2019excavating}. ML researchers reported that most benchmark datasets used in ML applications include widespread label inaccuracy \cite{Northcutt2021-wy}. Many efforts have been made to release benchmark datasets that relieve the issue of spurious correlations in the data \cite{Hu2020-rg}, datasets safe from social bias \cite{Yang2019-ix}, and to develop methods to obtain fine-grained data annotations that better align with the motivating task of the benchmark \cite{Tsipras_undated-sl}. On the other hand, recent findings further emphasize the need for well-made datasets. For instance, small models trained on high-quality small-size data often did better than large models trained on bad or unsuitable data sets \cite{Northcutt2021-wy}. 

Data-centric AI (DCAI) \cite{NgAI, Ngcompetition, DeepLearningAI2021-gt}, one recent trend in ML, calls for a transition of focus from models to data to improve ML systems. DCAI is the discipline of "systematically engineering the data used to build an AI system." It investigates what makes data "good" and how to systematize best practices for constructing datasets to enable novices and experts to continue producing quality data \cite{noauthor_undated-qk, Eyuboglu2022-jy}. In that regard, researchers investigated ways to complement and refine existing data while building new datasets that overcome previously stated challenges and suggested novel frameworks that enable more rigorous dataset development \cite{bloom-building, Mazumder2022-kg}. Many researchers also strived to provide guidelines and establish data-focused benchmarks. Gebru et al. \cite{Gebru2018-xt} introduced Datasheets for datasets, an extensive checklist composed of questions that dataset creators should try to answer. Covering the overall process of data creation, distribution, and maintenance, the datasheet aims to help data creators provide comprehensive documentation in favor of transparency and accountability while allowing data consumers to make informed choices when using the data. In addition, DataPerf \cite{Mazumder2022-kg} is a growing suite of benchmarks related to building, maintaining, and evaluating datasets to make such practices easier, cheaper, and more repeatable. These efforts combine to reorganize data development that happened in an unprincipled manner with post hoc rationales into more disciplined engineering practice. Researchers have provided a fuller landscape of data practices to facilitate a better understanding of the current state. However, there is still a lack of focus on humans, who imbue meaning into data to make it useful. To create, maintain, and use data, it always takes more than one person and different kinds of work.

\subsection{Data and HCI}

From an interview-based study with 16 ML practitioners, Xin et al. \cite{Xin2021} found that participants spent over half of their entire time in the ML pipeline on data preparation, with around 80\% of tasks in the stage performed manually. Numerous ML papers \cite{roh2021sample, Roh2018-al, Karmaker_Santu2021-oz} focused on data engineering in ML workflows and primarily introduced the technical skills required for specific steps in data work (e.g., data collection, cleansing, and validation). Little is known about how humans intervene, what they do, and what considerations they take in the data work across the ML pipeline.

Humans play a significant role in determining whether and how a dataset might be useful for a specific application, conducting thorough error analysis, and handling the numerous unstandardized difficulties of creating useful datasets. In that regard, understanding the human labor, human judgments and biases, and volatile circumstances involved in producing datasets for ML systems has emerged as a big issue \cite{Paullada2021-ja}.

Much work \cite{Zhang2020How, Cartwright2019, Anik2021Data, Wang2022whose, Thakkar2022when} in the field of HCI is bridging the gap between investigating technical skills for building datasets and understanding humans' roles and influences in making ML data. Studies have also covered the organization and infrastructure surrounding annotation. Wang et al. \cite{Wang2022whose} suggest a view of data annotation as organized employment. Researchers have found that systematic work practices and well-organized annotation processes help annotation businesses and requesters more than they help the workers.

Moreover, existing studies have focused on the annotators who actually build the data, investigating their work practices, and urging the need to acknowledge their career goals and aspirations. Rivera et al. \cite{Rivera2021-vp} figured out the career goals and challenges of crowd workers on Amazon Mechanical Turk. Participants in AMT desired to pursue a profession other than crowd labor and signed up for AMT as a starting point, but many struggled to take further steps owing to a lack of career advice and constrained time and financial resources. In addition, Zhang et al. \cite{Zhang2022-ga} noted that developing annotation tools is time-consuming, costly, and requires software expertise. To alleviate such burdens, they suggested a conceptual framework to help the easy development of annotation tools by considering multiple usage cases. Based on the framework, they proposed the OneLabeler, designed with reusability and flexibility in mind. To sum up, previous literature has mainly focused on illustrating particular stakeholders' needs \cite{Cartwright2019, Rivera2021-vp, Wang2022whose} or suggesting systems to support humans in specific steps. 

Recently, the CHI 2022 workshop \cite{Pine2022investigating}, "Investigating Data Work Across Domains: New Perspectives on the Work of Creating Data," called for attention to investigate the details of data work to realize its goal of making data useful and meaningful. They highlighted the need to understand how humans create, collect, manage, curate, analyze, interpret, and communicate data. In this vein, we focused on the roles humans play, the difficulties, and the efforts to create good data in the data practices in ML workflows. 

\section{Method}

This paper seeks to understand what ML experts go through as they build and manage datasets to develop ML models. With ML practitioners in academia and industry working on various ML tasks, we conducted one-on-one semi-structured interviews, with a task to build a dataset construction pipeline as supplementary material alongside the interview. Participants were asked about their practices, challenges, and support they think are necessary to construct datasets for ML system development.

\begin{table}[!t]
\centering
\renewcommand{\arraystretch}{1.2}
\caption{Participant demographics, from the left: (1) Participant \#, (2) Job/role, (3) Years of experience in programming + ML, (4) Domain, (5) Workshop task.}
\label{tab:study participants}
\begin{tabular}{@{}llcll@{}}
\toprule
No. & Job/Role & Yrs. of Exp. & Domain & Workshop Task \\ \midrule
P1 & AI Research Engineer & 6 & NLP & Opinion mining from product reviews \\
P2 & AI Research Engineer & 5 & CV, Multimodal & Physical activity recognition \\
P3 & ML Engineer & 7 & CV, OCR & Checkbox recognition from documents \\
P4 & Data Manager & 2 & CV, OCR & Medical documents parsing \\
P5 & AI Research Engineer & 8 & CV & Detecting defects in manufactured products \\
P6 & AI Research Engineer & 4 & NLP, Speech & Automatic speech recognition \\
P7 & AI Research Engineer & 11 & OCR & Medical documents parsing \\
P8 & AI Research Engineer & 7.5 & CV, Bioinformatics & Detecting defects in manufactured products \\
P9 & AI Research Engineer & 4 & CV, Medical & Detecting tables in documents \\
P10 & Software Engineer & 13 & NLP & Neural Machine Translation \\
P11 & Data Scientist & 7 & CV & Fashion image search and recommendation \\
P12 & AI Research Engineer & 6 & CV & Defect detection for mobile phones \\
P13 & Ph.D. Student & 9 & MIR & Automatic DJ/mixing \\
P14 & Software Engineer & 3 & OCR, CV & Credit card OCR \\
P15 & Ph.D. Student & 4 & Speech, MIR & Singing Voice Synthesis \\
P16 & Ph.D. Student & 8 & NLP & Neural Machine Translation \\
P17 & Ph.D. Student & 5 & CV & Object detection for classifying mart goods \\
P18 & AI Research Engineer & 3 & NLP & Hate speech detection \\
P19 & AI Research Engineer & 2 & NLP & Sentiment analysis from movie reviews \\ \bottomrule
\end{tabular}
\end{table}

\subsection{Recruitment \& Participants}

Participants were recruited from the authors’ personal networks, developer communities, and school communities via snowball and purposive sampling. As a result, 19 participants affiliated with 13 different organizations were recruited. In particular, we recruited engineers and researchers who have experience with the following: 

\begin{enumerate}
  \item Constructing datasets for supervised learning and labeling work associated with.
  \item Developing ML models for real-world applications.
  \item Using AutoML tools in practice (or having an understanding of AutoML).
\end{enumerate}

Overall, our participants had a minimum of 2 to a maximum of 13 years of experience with ML and programming. All participants had experience developing ML models for real-world applications or conducting research projects. Their profile spanned multiple occupations, ML domains/applications, and affiliations. Table \ref{tab:study participants} summarizes participant demographics, including their years of experience in programming + ML, current roles, and domains of ML they have expertise in. The last column shows specific tasks each participant chose for the workshop task. The details of participants’ affiliations are omitted from the table for anonymity. Participants were compensated for each interview with about 50,000 KRW (approx. \$40).

\subsection{Study Procedure}
Our participants are located in or have worked on projects based in South Korea or the USA. We conducted all interviews in Korean, the first language of participants and interviewers. All interviews were conducted online via Zoom, considering COVID-19 situations. 

The interview began with researchers briefly introducing the research goal and asking participants about their backgrounds and experiences in constructing a dataset for developing an ML application. Participants were then asked to expand upon the experience they had just described and were tasked with building a data construction pipeline. As a prompt for the task, participants were asked the following questions:

\begin{figure}
    \centering
    \includegraphics[width=\textwidth]{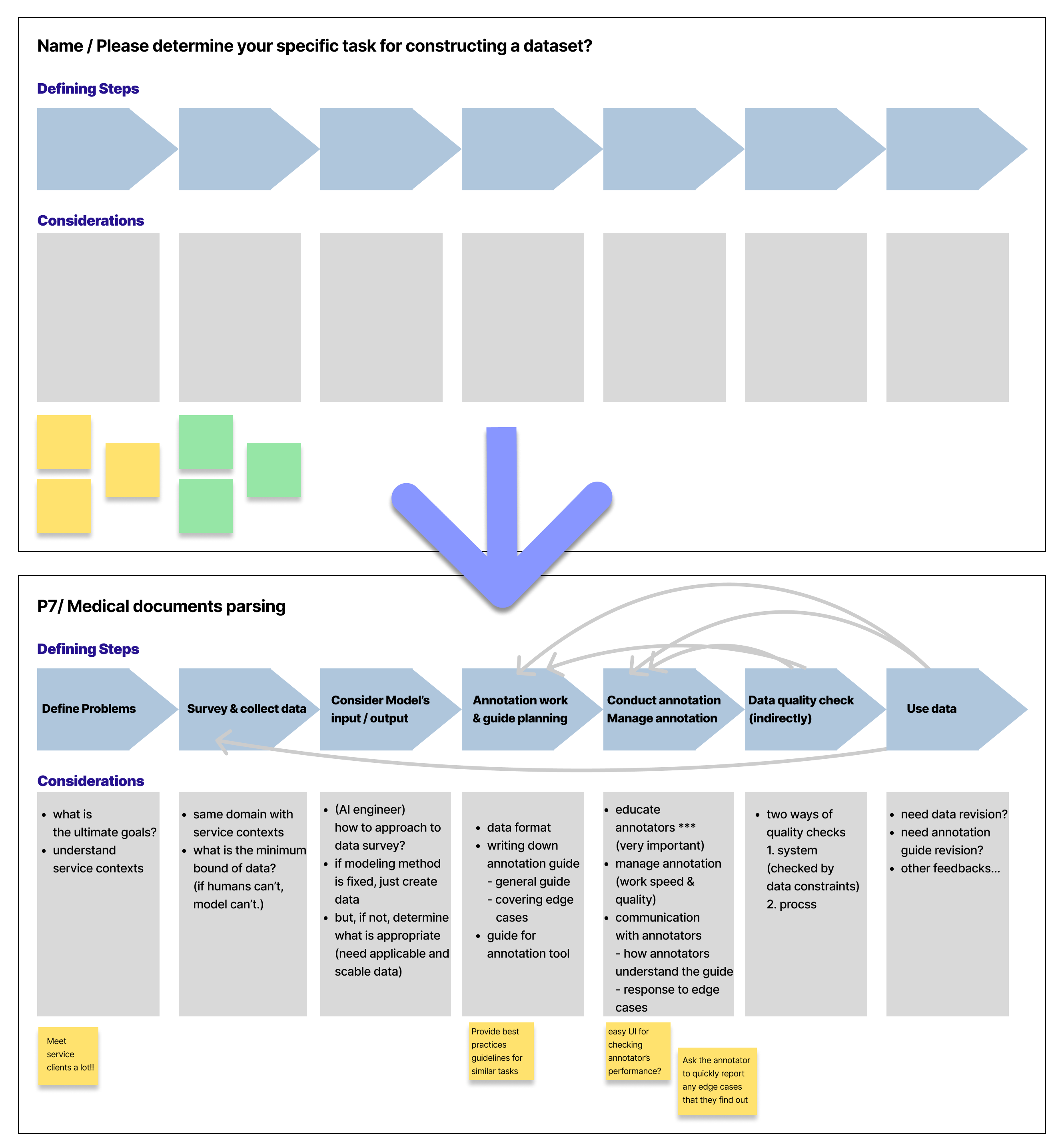}
    \caption{A screenshot of a Figjam worksheet, where top: barebone, and bottom: an example of a filled-out worksheet. Participants were instructed to enter the steps of data construction at a higher level in the upper blue boxes and specific tasks and considerations for each step in the gray boxes below. As they proceeded, participants used post-its to add any additional details.}
    \label{fig:figure1}
\end{figure}

\begin{itemize}
  \item \textbf{Assuming constructing a dataset to develop an ML model to deploy for real-world application, what pipeline do you go through?} What are the steps in the pipeline? What tasks are performed at each step? What specific challenges and considerations need to be addressed in each step?
\end{itemize}

To assist participants in elaborating upon their experiences, we provided a worksheet, as in Figure.\ref{fig:figure1}, and asked them to fill in the barebone pipeline with steps of data construction, along with detailed challenges and considerations for constructing good data in each step.

Participants were asked to think aloud about their experiences and opinions fully throughout the task. Participants explained the process of constructing a dataset in a specific domain/task of their choice (see “Workshop Task” column in Table \ref{tab:study participants}) to deploy an ML model trained with the data to the general public. Additionally, participants were instructed to choose a target task based on constructing unstructured data for supervised learning since human involvement is more natural for unstructured data.

Upon completing a data pipeline, the participants were then asked another set of questions that intend to investigate human engagement in data construction as it is a task that inherently involves human-model interaction:

\begin{itemize}
  \item \textbf{What roles do humans play in data construction?} What parts must humans intervene in and use their tacit knowledge? How do they interact with data and models to make good data in the process? What parts could humans benefit from interacting with models, i.e., Would gaining any information from the model help construct better data? 
\end{itemize}

Finally, we also asked the participants what and how they would support non-expert users to build datasets by following the data construction pipelines.

\begin{itemize}
  \item \textbf{If non-experts were to create datasets following the pipeline, what help would they need to minimize trial and error?} Which steps in data construction could be automated to help non-experts easily construct datasets? Which steps would be impossible to automate? Are there parts that mandate human intervention? Why?
\end{itemize}

Note that participants were given a specific context of helping non-experts create a dataset, as we set to provide them a window to tap into challenges that they think are the most critical yet systematically solvable problems when constructing data from a self-distanced perspective.

\subsection{Analysis \& Coding}

All interview sessions were video recorded with participants’ consent, resulting in 30+ hours of recording. Then from participants’ data construction pipelines and interview transcriptions, about 90 action items for constructing a dataset were extracted with overlaps. Three members of our research team reorganized the participants’ responses and clustered similar action items together under 6 major steps, which we derived from previous works \cite{Karmaker_Santu2021-oz, Roh2018-al, Whang2020-du} that lay out generalized steps to construct a dataset for developing an ML system, to construct a generalized data construction pipeline with 14 unique action items. We also organized considerations and challenges in each step of data construction. This analysis was performed iteratively altogether.

Further, we analyzed participants’ interview transcriptions in depth to deduce insights for developing good data. This thematic analysis was done using Dovetail \cite{dovetail}, an online UX research/markup tool. Similar to the affinity diagramming method, each researcher individually highlighted noteworthy quotes from participants and assigned descriptive codes (such as “difficulties in data construction” or “requires domain knowledge”  to encode what piece information a quote contains or an insight it provides. After the individual analysis, the researchers came together for a collaborative deductive coding process, and 284 codes in total were extracted. Our analysis team categorized these 284 initial codes into 39 sub-level themes and further grouped them into three major themes.

\section{Findings}

This section presents our findings from interviews with 19 ML experts and the thematic analysis of interview responses. Through our findings, taking a step beyond painting a broad landscape of data construction work, we scrutinize people's realistic considerations and dilemmas in making good data for ML. This section includes two parts: 1) what humans do and what they consider in each step of the data construction pipeline, and 2) what are recurring struggles and endeavors in the steps of constructing good data. 

\begin{figure}[!t]
    \centering
    \includegraphics[width=\textwidth]{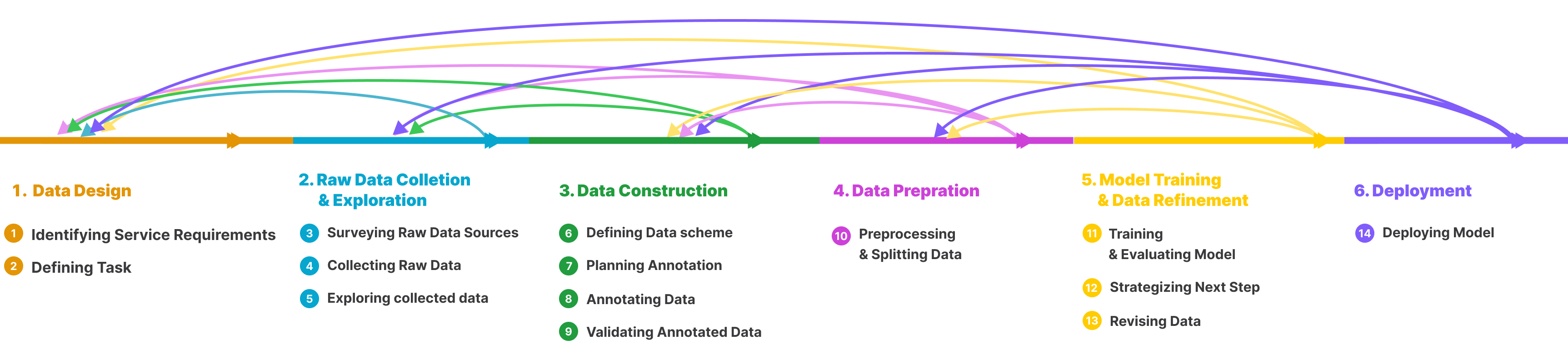}
    \caption{Data Construction Pipeline}
    \label{fig:figure2}
\end{figure}

\subsection{What do humans do and consider in the dataset construction pipeline?}

As described in section 3.3, we reorganized our 19 interview participants’ data construction pipelines into one generalized dataset construction pipeline, as shown in Fig \ref{fig:figure2}. Table \ref{tab:Task} presents that the pipeline includes six steps, each composed of 1 to 4 sub-steps. In the following, we elaborate upon the pipeline, describing what humans do and consider when they make data.

\subsubsection{Step 1 – Data Design}


\begin{enumerate} \addtocounter{enumi}{0}
  \item \textbf{Identifying service requirements – [service values and operational contexts]} 
  
     The characteristics of the desired ML services/products and their goals were identified for the first step of data design. Participants who have built ML models for service stressed the importance of understanding how models will be used in service. Knowing what values an ML-driven service intends to deliver [P1, P7, P12], what demands prospective service users may have, and in what environment a model would operate [P11] were essential for planning ahead how data should be looked at. P17, with his experience of developing a product classification model to be deployed in an autonomous robot inside a supermarket, said
     
     \begin{quote}
         “Understanding the store environment was essential for determining how a model should operate as all stores have different arrangements of products and layouts, and even the amount of products displayed on shelves is different, while all of that may influence a model’s working.” (P17)
     \end{quote}
\end{enumerate}

\begin{enumerate} \addtocounter{enumi}{1}
  \item \textbf{Defining the task – [target domains, task difficulty and formulation, pilot study, literature survey]} 

A step following identifying the service requirements was defining what problem a model solves. According to P14, it was necessary to assess whether a conventional rule-driven approach is sufficient or a data-driven approach is required to develop an algorithm for a service, as the latter requires data collection that can be costly and time-consuming. If the solution required data, then estimating the minimum/optimum amount of required data came next [P4, P7], and determining the problem formulation [P8] and the model architecture and training methods [P3, P7] were subsequent steps. Referring to how others previously approached similar problems could also be effective in determining the amount of data required and setting a model’s objective.
\end{enumerate}

\begin{table}[!ht]
\centering
\caption{Definition and Tasks in Each Step of Data Construction Pipeline.}
\label{tab:Task}
\resizebox{\textwidth}{!}{%
\begin{tabular}{@{}lll@{}}
\toprule
Step & \textbf{Sub-step} & \textbf{Definition \& Tasks} \\ \midrule
\multirow{2}{*}{\textbf{1. Data Design}} & (1) Identifying Service Requirements & - Identify characteristics of the desired AI service(product) and understand its goals \\ \cmidrule(l){2-3} 
 & (2) Defining Task & \begin{tabular}[c]{@{}l@{}}Define the model's objectives\\ - Determine model I/O\\ - Determine methodology\\ - Survey previous works\end{tabular} \\ \midrule
\multirow{3}{*}{\textbf{2. Raw Data Collection and Exploration}} & (3) Surveying Raw Data Sources & \begin{tabular}[c]{@{}l@{}}Determine raw data sources\\    - List-up and explore raw data sources\end{tabular} \\ \cmidrule(l){2-3} 
 & (4) Collecting Raw Data & \begin{tabular}[c]{@{}l@{}}Data collection\\ - Crawl or purchase existing data\\ - Create new data\end{tabular} \\ \cmidrule(l){2-3} 
 & (5) Exploring Collected Data & Exploratory data analysis \& visual inspection \\ \midrule
\multirow{4}{*}{\textbf{3. Data Construction}} & (6) Define Data Scheme & \begin{tabular}[c]{@{}l@{}}- Determine model evaluation metric (e.g., accuracy, F-score)\\ - Define annotation scheme\\    - Annotation format\\    - Annotation method\\    - Label hierarchy, etc.\end{tabular} \\ \cmidrule(l){2-3} 
 & (7) Planning Annotation & \begin{tabular}[c]{@{}l@{}}- Define annotation guideline\\ - Select annotation tool\\ - Recruit annotators or hire data collection agency\\ - Evaluate annotators' knowledge, skills, and experiences\end{tabular} \\ \cmidrule(l){2-3} 
 & (8) Annotating Data & \begin{tabular}[c]{@{}l@{}}- Educate annotators\\ - Auto-labeling\\ - Annotate raw data\\ - Manage annotators\\ - Q\&A for annotators\end{tabular} \\ \cmidrule(l){2-3} 
 & (9) Validating Annotated Data & \begin{tabular}[c]{@{}l@{}}- Cross-validation among annotators\\ - Qualitative evaluation of annotations\end{tabular} \\ \midrule
\textbf{4. Data Preparation} & (10) Preprocessing \& Splitting Data & \begin{tabular}[c]{@{}l@{}}- Split data to train/test/valid sets\\ - Preprocess data to match the format of model input\end{tabular} \\ \midrule
\multirow{3}{*}{\textbf{5. Model Training and Evaluation}} & (11) Training and Evaluating Model & \begin{tabular}[c]{@{}l@{}}- Train baseline model(s) \\ - Select 2-3 different models and train/evaluate with the same data\\ - Estimate model performance with small data\\ - Qualitative \& quantitative evaluation of model \\ - Visualization\end{tabular} \\ \cmidrule(l){2-3} 
 & (12) Strategizing Next Step & \begin{tabular}[c]{@{}l@{}}- Identify samples model didn't perform well \\ - Determine the source of problem \\ - Model?  Data?  Annotation?\end{tabular} \\ \cmidrule(l){2-3} 
 & (13) Revising Data & \begin{tabular}[c]{@{}l@{}}- Change model OR add more data OR change annotation guide\\ - Version new data\end{tabular} \\ \midrule
\textbf{6. Deployment} & (14) Deploying Model & \begin{tabular}[c]{@{}l@{}}- Deploy for service\\ - Continued maintenance\end{tabular} \\ \bottomrule
\end{tabular}%
}
\end{table}

\subsubsection{Step 2 – Raw Data Collection and Exploration}

\begin{enumerate}\addtocounter{enumi}{2}
  \item \textbf{Surveying raw data sources – [source, availability, representativeness, specifications, legal and ethical issues]} 
  
  Surveying raw data sources, selecting appropriate sources, and deciding methods for data collection were the main tasks of this step. It was necessary to estimate the amount of data feasible for collection [P4, P5, P9, P10, P14] and select the types of data to be included or excluded [P7]. P5 mentioned that it is vital to consider the detailed characteristics of data, such as image resolution and sampling rates for audios, and what other data were in use in the target domain. Potential legal (e.g., copyrights) and ethical issues with data must also be checked [P4, P13, P16, P19]. In particular, P19 put extra care not to introduce human bias during data collection, as it was easy to blindly attempt to collect more data without careful consideration of data sources when more data often results in a better performance. P19 also noted:
  
     \begin{quote}
“If I were to only crawl data from the website with the most community traffic in Korea (DCInside), because it has the most data, and train a language model. That model would likely learn explicit insults and even sexually discriminative comments. (...) That means there is a discrepancy between how the real-world data looks and what people want. Possibly that is the nature of the Internet, and that is why I was interested in cleaning corpuses crawled from the web. As more data usually results in a better model.”
     \end{quote}
     
P4 also emphasized that the data must represent the entire domain, especially when a sampled subset of data is used.

\end{enumerate}

\begin{enumerate} \addtocounter{enumi}{3}
    \item \textbf{Collecting Raw Data – [quantity, cost, quality, privacy, speed, validation]} 
  Data were collected from existing data or entirely anew from scratch. The cost of data collection in terms of both time and money must be put into consideration [P4, P13, P14]. Collected data were then checked for duplicates and any errors, such as missing values or broken links [P2, P10, P13, P16], and either rule-based or model-based methods were used for cleaning error. It was also necessary to decide how to deal with privacy-sensitive data (if there is any), for example to delete them or apply masking [P4]. Additionally, P2 and P13 noted it is important not to overload websites when using web crawlers to collect data. 
\end{enumerate}

\begin{enumerate} \addtocounter{enumi}{4}
  \item \textbf{Exploring collected data – [class distribution, noise, manual inspection]} 
  
  Collected data must be explored and, if necessary, validated with bare eyes. Many of our participants suggested it is crucial to check for noises in data and class distribution. Removing dirty data could provide more significant benefits than collecting more data, and often a small portion of contaminated data detrimentally harms the overall quality; thus, what constitutes noisy data has to be defined carefully. P5 explained that when detecting defects from electronic products, the various types of defects and backgrounds that are not defects were identified from images. In that process, all data were inspected manually if data was sufficiently small; otherwise, only a subset. The participant also noted that having “defective” data was as important as having “normal” data, as balanced data is essential for a well-performing model.
\end{enumerate}

\subsubsection{Step 3 – Data Construction}
\begin{enumerate} \addtocounter{enumi}{5}
  \item \textbf{Defining data scheme - [evaluation metric, annotation method, test set, alignment with model]} 
  
  The annotation schema, including data formats, label categories, and annotation methods, were defined along with the evaluation metric (e.g., accuracy and F1-score) in this step. Surveying schemas and metrics used in other similar datasets came first [P1, P14]. The scalability of data had to be taken into account when selecting an appropriate annotation method to use.
  
     \begin{quote}
“For example in OCR, it is possible to use either center points or bounding boxes when annotating texts, and while bounding boxes are more costly than center points, they are more scalable as center points can be easily computed from the points of a bounding box.” (P3)
     \end{quote}
     
     P1 emphasized that labeling categories should be constructed, so labels are mutually exclusive but collectively exhaustive. Evaluation metrics must be considered along with labels and annotations, i.e., whether a metric matches a model’s performance as experienced by the service users [P12, P14]. P5 shared an experience where the model’s performance metric and service goal had to be closely aligned.

     \begin{quote}
         “We had to take into account what our users expected. When a defect detection model makes an error, it can be either a false positive (okay but classified as defect) or a false negative (defect but classified as okay). Because FN is more critical than FP when it comes to detecting defects, we put more focus on minimizing the FN and tested our models towards that goal.” (P5)

     \end{quote}
     
     Some found that formulating a test set before building other parts of a dataset was also critical for constructing good data. P4 and P10 noted that a test set should closely represent the real world, while P8 and P15 also noted that a test set must include samples representative of a model’s training.
     
\end{enumerate}

\begin{enumerate} \addtocounter{enumi}{6}
  \item \textbf{Planning annotation - [annotation cost/speed/quality, hiring annotators, consistency of an annotation guide, annotation tool, pilot study]
}

How raw data should be annotated was determined in this step. Hire annotators and let them work in-house, outsource tasks to a crowdsourcing agency such as MTurk, or do it manually? This decision required considering time, money, and expected quality of data. Once the annotation method was determined, an appropriate tool was surveyed and chosen or implemented from scratch if no existing tool fit the purpose.

The following steps were writing an annotation guideline and determining qualifying criteria for hiring annotators. Our participants found careful planning is mandatory as annotation tasks are “expensive” with constraints in both time and money. The number of annotators was chosen from the time available for the task [P4, P9, P14, P16, P18] and the available time in turn was determined from the allocated budget [P4, P14, P16].

Consistency was of utmost importance for annotation guidelines to minimize any chance annotators produce disparate annotations when given the same data [P1, P3, P4, P6]. Educating annotators such that they well understand the guideline was another point of consideration [P4, P5, P16, P19].

Finding a suitable annotation tool for the task was necessary as data requirements are heterogeneous across tasks. P9 and P14 reported implementing an annotation tool from scratch to meet the requirements of their project, as no existing tool satisfied their needs. In common, our participants reported that an annotation tool’s supportive features directly influence the quality of resulting data, such as the functionality to analyze annotation logs, manage the performance of annotators, or even simple shortcuts that help increase productivity.

Naturally, the importance of who builds data was highlighted as well. Participants noted that the qualification criteria for annotators have to be set by the specifics of the task and data [P5, P8, P9, P14, P15], and not only the backgrounds of annotators but the backgrounds of task managers writing guidelines and managing the workers must be considered to mitigate potential biases in data [P18].

P18 mentioned that getting hands dirty in the annotation tasks also helps to understand the potential challenges for annotators and what might be done in their support. P19 similarly noted that constructing a small pilot dataset before getting into the main task helps foresee potential data construction problems, which can be reflected in an annotation guideline and the overall plan ahead.

\end{enumerate}

\begin{enumerate} \addtocounter{enumi}{7}
  \item \textbf{Annotating data - [annotator well-being \& conditions, edge cases, annotation guide update]} 
  
  Aside from the annotation task itself, managing and educating the annotators, doing Q\&A for annotations, and automating data annotation with the assist of a model are included in this step.
 
Following the education with annotation guidelines in the previous step, workers were educated on performing annotation [P4, P11, P19]. P4 also envisioned annotators how the data they annotate would be used in models and for what services, which helped workers be more motivated and productive as a result. P19 also found showing various examples of images the workers will encounter during annotation was far more effective than explaining the guideline only in words to increase the consistency of annotations. Almost all participants agreed that annotators themselves largely decide the quality of data, and some also mentioned that managing the annotation workload, not to fatigue workers is important for both the efficiency and consistency of annotations [P4, P7, P11, P14, P18, P19].
 
P4, P7, and P9 shared workers often submit previously unreported edge cases during the annotation, and merging new cases into the guideline should be done with the care that there are no conflicts with completed annotations.
 
Further, if available, a trained model can be used for automated annotation; tasks can be sped up significantly by human annotators revising automated annotations compared to manually annotated data from scratch.

\end{enumerate}

\begin{enumerate} \addtocounter{enumi}{8}
  \item \textbf{Validating annotated data - [inter-annotator agreement, manual validation]} 
  
  The quality of annotated data is evaluated in this step. Participants cross-validate annotations across multiple workers, or assessed annotation results qualitatively (i.e., manually). P19 measured inter-annotator agreement across multiple workers, while P11 had a well-performing annotator (or engineers themselves) manually inspect annotators’ works.

\end{enumerate}

\subsubsection{Step 4 – Data Preparation}

\begin{enumerate} \addtocounter{enumi}{9}
  \item \textbf{Preprocessing and splitting data – [memory constraints, split strategy, test set quality]} 
  
  A complete dataset is constructed with annotated \& validated data in this step. The dataset is divided into train, validation, and test splits and preprocessed to match the model’s input format. P8 emphasized stratifying splits, while P12 and P15 noted that a test split should include diversified samples. Random sampling can be sufficient to obtain balanced distribution across splits if the data quantity is large [P12, P15], but the data distribution tends not to follow actual data distribution, especially in the early stage of data construction when the data size is small. When the data has imbalanced distribution across classes, extra care should be taken that each split includes at least one instance of every class [P5]. Additionally, the resulting data should be saved in a drive so that they are easily accessible [P9], along with versioning and memory constraints also under consideration [P3].

\end{enumerate}

\subsubsection{Step 5 – Model Training and Data Refinement}

\begin{enumerate} \addtocounter{enumi}{10}
  \item \textbf{Training and evaluating model – [wrong label, service requirements]} 
  
A model is trained and evaluated in this step. Participants reported training and evaluating multiplied models in parallel on the same data [P8, P17, P18, P19] and selecting a subset of samples for a quick test run before training a model on the full dataset [P3, P5].
 
For model evaluation, the test scores were put first [P18], along with other measures such as the inference speed and confidence scores [P1, P11, P14, and P15]. Qualitative evaluation was also performed; for example, P5, P11, and P16 searched for cases where models’ prediction and ground truth labels disagree, as in some cases, the annotated labels were wrong, but the prediction was correct. It was also brought up that evaluating the model in service context was crucial, prioritizing the service requirements over evaluation metrics [P5].

\end{enumerate}

\begin{enumerate} \addtocounter{enumi}{11}
  \item \textbf{Strategizing next step – [quantity, edge cases, data quantity, model refinement, data refinement/debugging]} 
 
  A decision is made to deploy a model or go through another round of training. This step requires understanding the characteristics of samples a model performs poorly and deciding whether a model or data should be fixed in response. If data is the cause, it is necessary to determine if increasing the data quantity is sufficient or if new samples are needed. P13 and P14 suggested additional data collection with a revised annotation guideline or preprocessing strategy unless refining a model is enough to fix the problem.

\end{enumerate}

\begin{enumerate}\addtocounter{enumi}{12}
  \item \textbf{ Revising data - [annotation guideline, difficult samples, synthesized data]}

Based on the decisions from the previous step, additional data is collected, the model is refined, or annotation guidelines are updated. If types of samples model perform poorly are known, new data of similar type can be collected [P1, P2, P4, P6, P7, P8, P9, P10]. Along with another round of training and evaluation, an annotation guideline is updated and the label categories are redefined if necessary [P1, P2, P3, P4, P5, P7, P15]. If possible, trained models can be utilized to generate synthetic data [P10]. Careful versioning is required as the data goes through changes during these iterations.

\end{enumerate}

\subsubsection{Step 6 – Deployment}

\begin{enumerate}  \addtocounter{enumi}{13}
  \item \textbf{Deploying model - [unseen data, re-collect]}
  
  At last, a model is deployed for service. From this point onward, continued maintenance of a model is necessary by keeping a model robust against incoming real-world data that is possibly out of the model's learned distribution. Planning for additional data construction and model training meanwhile inspecting incoming service data may be necessary [P16, P18].
\end{enumerate}

All participants mentioned that data construction is an iterative process. They note that the iteration does not happen only after evaluating a model, but in any step of the pipeline; data construction is not a linear unidirectional operation but a bidirectional, cyclic process.

\subsection{What challenges exist towards making good data? How to deal with them?}
In this section, we focus on the overarching themes that repeat across each step in the pipeline. Such themes are best described as the inevitable struggles toward making good data. We present the difficulties the participants face when making data for ML and their strategies to deal with these struggles.

\subsubsection{Trade-offs stemming from real-world constraints}
P14, P16, and P10 reported that data construction is costly. Therefore, making good data by following the ML and data management disciplines in practice is difficult because of the realistic constraints, such as time and cost. 
 
The participants considered the tradeoff between short-term and long-term costs in designing data. Specifically, P1, P14, and P19 argued the importance of scalable data design to respond to different models and domains. They mentioned that if they had planned annotations with multiple options to prepare for situations that can be used on diverse models, they would have to pay more costs in the short term. However, if they wanted to reduce short-term costs, they could make data quickly, but the data would not be scalable.

\begin{quote}
    “If we want to use our data more than just once, continually adapting its format for various purposes, we will need to make sure the currently defined format is scalable. It won’t be easy, but surely it will help cut cost if we were to go large in quantity” (P19)
\end{quote}

The participants shared how they dealt with such dilemmas with their own strategies. P18, P10, and P11 emphasized quick evaluation through constructing a pilot dataset for long-term cost reduction. To be specific, P10 started building data after determining whether the performance would be improved if data was added through quick evaluation using public datasets. P18 conducted pilot annotations to identify problems or edge cases that may appear during this construction in advance. He said that this pilot annotation minimizes trial errors when building data. P11 conducted data debugging after examining the output of the model trained with small sample data.

The participants also mentioned trade-offs when managing the annotators and preparing guidelines for annotation. P6 and P17 reported the trade-off between speed and validity \citep{Aroyo2022-tx} of data. Sometimes the annotation schema that accurately represents the phenomena to capture might be very complex. Such schemas may impose too much cognitive load on annotators, slowing down the speed of data annotation. P17 shared his experience of reducing the number of label types to speed up data production at the expense of data validity. In addition, P4 considered balance in validity and consistency when writing annotation guidelines. Giving specific directions on annotating ambiguous cases may help increase consistency among annotators; however, unifying the label for all subjective cases may negatively affect the validity of data. 
 
The participants have established strategies to deal with such dilemmas in managing annotators and guidelines. After the first dataset construction, P16 set strategies to identify the quantity of data for additional construction for saturation points. P15 and P17 said they revised data (e.g., annotation schema, annotation guideline) based on model performance. Furthermore, P19 stressed the importance of the dataset being easily reused. To do so, data managers and modelers wrote data documentation and their specifications with problem definition, data format, and baseline model performance.
 
The participants collectively mentioned various types of trade-offs in data work. Such dilemmas resulted from cost, scalability, data validity and consistency. Nevertheless, participants took strategies to pursue data excellence, that is, to spend cost efficiently without abandoning data quality. 

\subsubsection{Harmonizing assorted data workers for consistency}

Our participants collectively state that bringing together the differences that stem from diverse people who participate in the data work is difficult. Everyone has a different point of view on the same data or has only partial knowledge required to make good data. P10 and P12 addressed that the results vary depending on the knowledge about the target service and model. It was difficult for P10 to determine whether the data met the service requirements due to a lack of domain knowledge. P17 noted that the modelers need domain experts' help when modeling to reduce the gap between service and model requirements.

\begin{quote}
     
“So we were making a translation service, a multilingual one. And we didn’t know all the languages, but we still made test sets and what not, all of that wasn’t so easy.” (P10)

\end{quote}

Participants said that the required level of knowledge was different depending on what role they played in the process of constructing data. For example, P16 said that a person who determines annotation schema or performs annotation does not need to understand the specifics of ML. However, it would be helpful for them to have basic knowledge of ML (e.g., ML systems work based on distribution, the model's input/output, and the current model's pros and cons). Likewise, P3 mentioned that it is very important to educate a person who makes data on how the model is evaluated or how the model works. P11 pointed out that knowing the service domain (e.g., whether there is a clear difference between casual and minimal) can help whether the model can accept such annotation structure can be determined in advance without trial and error. 

During the annotation, the quality and quantity of work may differ significantly depending on the knowledge level and the annotator's well-being and motivation. P6 emphasizes that in some tasks where more data guarantees higher performance (e.g., speech-to-text model),\textit{"It's all about the caliber of annotators,"} such as how good they are at transcription and how fast they type, has more influence on the system performance than any other characteristics (e.g., data distribution or data sources). 
 
The participants endeavored not to cause bias in the model and data-driven by annotators' different knowledge levels and biases. In particular, these characteristics affect each step of annotation guides and task management, which are the most critical factors in maintaining data quality. In defining the task and planning annotations, P6 said several people should write the guide to prevent author bias. P1, P3, and P12 use examples to prevent arbitrary interpretation of the guide and to show clear intent. P4 and P7 said that just giving guidelines is insufficient to make good data. Therefore, they used various methods of worker training, such as showing animations or videos of sample works. 

\begin{quote}
    “Overall, the workers follow a similar flow, but their details can be different. Just with a guide, you can’t convey what you think should exactly happen in their mind as they make data. But when you educate the workers, you first give them data and show the sequence of how your attention flows through as you work through data, be it a video recording, then the workers will flow your flow. This boosts the consistency of data and the types of edge cases the workers report become similar as well.” (P7)
\end{quote}

Participants specifically attempted to reduce these differences through direct communication at the step of annotating data. P3, P6, P7, and P15 suggested that the modeler or data manager communicates with the annotator constantly and on time to reduce the gap in guide understanding. In addition, P12 said that it is necessary to understand the appropriate amount of work for each annotator rather than assigning work as much as possible. These attempts aimed to maintain annotation quality consistently. After annotation, P12 reported that he checks the correlation of annotation among different workers to check if there is a noise stemming from certain annotators. This process facilitates the data cleansing process. 
 
From judging the dataset’s suitability for the task to managing the consistency among annotators, the participants pointed out the inevitable challenges they face due to human involvement in data construction. The strategies for making consistent, the high-quality dataset included communicating with domain experts to align the service requirements into ML tasks, putting much effort into educating annotators, and checking inter-annotator agreement. 

\subsubsection{The necessity of human intuition and tacit knowledge gained through repetitive experience}
 
Participants said that following the textbook knowledge of technical skills and the disciplines for making good data is not enough. P10, P16, and P17 emphasized the importance of knowledge gained through experience looking through much data. Specifically, P14 and P17 mentioned that looking into models’ outputs often is a must to find the parts where the model gets confused and revise the label categories to resolve the confusion.

\begin{quote}
    “It all comes down to an experience when we try to sort out where a model gets confused, and figuring those pain points in the first place. Who has gone through a lot has gut feelings about how things might go down when training a model, and what to do to get a model working straight.” (P14)
\end{quote}

Participants also said that many steps in data construction require thorough manual inspection. Human intervention is mainly involved in the steps related to evaluating the model performance. P17 claimed that quantitative metrics (e.g., accuracy) do not tell much about what exactly the system is good at, so qualitative evaluation of the output is a must. P1 also emphasized manually going through model predictions and model calibration since a high confidence score of the model does not guarantee the correct answer. Constructing test sets requires human intuition as well. P14 often split train, valid, test set manually, image by image, so that test set can represent the cases to be evaluated with high fidelity. P1, P6, and P10 stressed that humans must inspect test datasets thoroughly. 

\begin{quote}
    “Honestly, I just pick a day, and sink some time into it. I have once gone through images one by one, before splitting train, validation, and test sets.” (P14)
\end{quote}
 
The fundamental reason behind utilizing human intuition and tacit knowledge in data construction lies in the lack of interpretability of deep-learning models (i.e., the blackbox model) and the messy nature of data. P17 described the trickiness in interpreting the effect of data improvement on system performance even for the experts. 

\begin{quote}
    “I’ve been going straight in, but knowing the causality between improvements in data and model’s performance would have helped training (...) I have had glimpses about when putting in some data would work out, but only if that could be more concrete.” (P17)
\end{quote}

P6, P11, P16, and P17 claimed the necessity of interpreting the model output into something meaningful to humans. They mentioned that there is an extra interpretation step upon examining the model output (e.g., label distribution, heatmap) to decide the next step for system improvement. For example, P6 and P11 mentioned Gradient-weighted Class Activation Mapping \citep{Selvaraju_2019}, which visualizes which feature of the data the model refers to when making predictions. After looking at the weight heatmap and model output, they can decide which data influences model prediction.
 
Participants pointed out that data work requires human input due to the labyrinthine nature of data. P10 said that quantifying or predicting the behavior of ML systems is difficult due to many variables stemming from data, unlike software engineering. Moreover,  P3, P6, and P10 pinpointed that it is impossible to be prepared for all the edge cases and variance of data in advance, as is the nature of real-world data. P1, P13 told not to trust the ground truth 100\%, since they are human artifacts after all.

\begin{quote}
     “What constitutes checkbox areas can differ and the check marks can differ too. You need an exorbitant amount of data to be able to cover all possible types of checkboxes.” (P3)
\end{quote}

As such, current data work is indubitably a human endeavor. Building good data requires human builders with tacit knowledge only acquirable from empirical experiences and their manual, tedious efforts. Humans being embedded in data so deeply makes systematizing data work a seemingly intractable problem without any efficient solution. Even so, as data work is already developing towards a disciplined engineering exercise for greater accountability and transparency of ML systems originating from data, together with the more efficient and iterable creation of data, what might be the following concrete steps to be taken toward that goal?

\section{Discussion}

Our findings show iteratively improving the ML system with multiple rounds of data collection and refinement is the status quo. The participants shared diverse strategies to make good data for AI systems, all formulated through unique individual experiences, rather than adapted from a pre-established best practice. Taking one extra step further from documenting and organizing the practices we observed, we discuss the future directions of research to help reduce the trial \& error in the data construction pipeline based on the experiences and desires of the ML practitioners. 

\subsection{There is no “skeleton key” to refining data.}

How exactly should data be refined to make a better ML system? At the end of the interview, P6 gave one last remark: \textbf{\textit{"How can we make good data? I have been pondering, to no avail. Really, foreseeing what data gives a good model is what we all dream of. "}} Our observations firmly echo the statement "no established metrics for defining high-quality data exist yet" \citep{Aroyo2022-tx}, highlighting the necessity for collective efforts on data quality assessment.

Nevertheless, our participants hinted at some hopeful yet concrete next steps toward systematizing data improvement. First, a model-independent metric for evaluating data quality needs to be developed. Assessing the fidelity of data (i.e., the degree to which the dataset represents reality) is an important topic to be explored \citep{Batini2009-wn, Aroyo2022-tx}. In that regard, P10 said that one common mistake in ML is judging the whole picture just by looking at sampled subsets of the data. Many of our participants found it crucial to check for class distribution of the data, and it would be nice to have a go-to method to find out the statistical landscape of the data. Another way to ensure model-independent data quality would be to align the qualitative and quantitative metrics. Most participants went through annotated data with their own eyes and evaluated data qualitatively; however, this kind of heuristic and qualitative evaluation makes it difficult to determine the causal link between data improvement and system performance improvement. It is high time for the data management community to develop interpretable measures to assess the data quality.

Second, it should be able to directly use model results for debugging data. Many participants improved data based on model confidence scores or the analysis of model output. For example, P6 and P11 used Gradient-weighted Class Activation Mapping \citep{Selvaraju_2019}, which visualizes the data-model interaction. However, the model confidence score is still not very reliable, and it is difficult to understand the model behavior, especially if model outputs are all but just a handful of numbers, so more research on model explainability is needed \cite{gunning2017explainable, Barredo_Arrieta2020-xl}. Studies toward providing more interpretable results that explain why models behave a certain way rather than just providing numbers will help improve by deciding the specifics of the data to collect and add.

\subsection{Annotation is at the core of data quality.}

Unsurprisingly, most of our participants found the annotation quality is the most important aspect of the data. In the same line, all of our participants reported that annotators themselves largely determine the data quality. Our participants strived to minimize inconsistency in annotations by writing well-designed annotation guidelines and devising effective methods for educating annotators. Therefore, all point to the direction to focus on the stakeholders involved in the data annotation process. Investigating how might the performance of annotators be enhanced would largely help increasing the quality of the data. 

For the annotators, a well-designed interface enabling efficient annotation should be developed. One work that emphasizes designing an annotation tool customizable for the task at hand is \citep{Zhang2022-ga}. Likewise, our participants emphasized the supportive functionalities of an annotation tool that can help annotators reduce errors systematically. Moreover, an annotator's understanding of the ultimate goal of the ML systems they contribute to, along with an overview of how models operate, can positively influence the annotation quality. For example, P4 envisioned for annotators how the data they annotate would be utilized in models and for what services, hence increasing worker motivation and productivity in turn. Another approach to aligning annotators’ motivation with the project goal is to provide appropriate education or guidance to workers. Educating basic ML knowledge may have a positive impact on data quality, as previously investigated \cite{Zhu2014-op, Batini2009-wn}.

For the managers of the annotation process, it is necessary to devise a way to facilitate the annotators' task management in terms of annotation speed and well-being. For example, some participants suggested that annotation tools should support monitoring annotation pace and tracking outliers. Moreover, constant and quick communication is essential to maintain the consistency of data annotation and for the annotators and managers/modelers to be on the same page even in times of high volatility in annotation guidelines. The current tendency of the ML community to abstract away the human workers, disregarding their contributions \citep{Sambasivan2022deskill, Irani2013-vr, Gray2019-xb}, is toxic both for the annotation workers and the annotation requestors.

\section{Conclusion \& Future Works}

There is no established best practice for data work in the ML community. Therefore, we must first find out and record how people make data, what efforts they make to make it “good,” and do best practices based on that. We conducted semi-structured interviews with 19 ML practitioners who had experience constructing datasets for supervised learning to understand data work practice and challenges. Based on the interviews, we present a data-focused pipeline for end-to-end ML systems, including the specifics of what humans do and consider for data construction. Our findings empirically show that data work requires a myriad of concerns involving a constant interplay of data, models, and humans. We identified recurrent human challenges at different stages of the data-building process: 1) trade-offs from real-world constraints; 2) harmonizing assorted data workers for consistency; 3) the necessity of human intuition and tacit knowledge for processing data. Based on the experiences and aspirations of the ML practitioners, we explored future research directions to decrease trial and error in the data-building pipeline. While the scope of our work was limited to making datasets for supervised learning techniques, we acknowledge that the degree or types of human intervention in the data construction process may differ in different situations, such as under semi-supervised learning or weak supervision circumstances.


\bibliographystyle{ACM-Reference-Format}
\bibliography{Reference}


\end{document}